\batchmode
\documentstyle[12pt]{article}
\newcommand{\nc}{\newcommand}
\nc{\renc}{\renewcommand}

\nc{\half}{{\textstyle{1\over2}}}
\nc{\etal}{\mbox{\it et al. }}
\nc{\ie}{{\it i.e.}}
\nc{\eg}{{\it e.g.}}

\renc{\thefootnote}{\arabic{footnote}}
\nc{\capt}[1]{{\bf Figure.} {\small\sl #1}}

\nc{\eqs}[2]{\mbox{Eqs.~(\ref{#1},\,\ref{#2})}}
\nc{\eq}[1]{\mbox{Eq.~(\ref{#1})}}

\nc{\figs}[2]{\mbox{Figs.~(\ref{#1},\,\ref{#2})}}
\nc{\fig}[1]{\mbox{Fig~.(\ref{#1})}}

\nc{\tag}[1]{\label{#1} \marginpar{{\footnotesize #1}}}
\nc{\mtag}[1]{\label{#1} \mbox{\marginpar{{\footnotesize #1}}}}
\renc{\baselinestretch}{1.5}
\newlength{\overeqskip}
\newlength{\undereqskip}
\nc{\be}[1]{\begin{equation} \mbox{$\label{#1}$}}
\nc{\bea}[1]{\begin{eqnarray} \mbox{$\label{#1}$}}
\nc{\Section}[2]{\section{#2}\label{#1}}
\nc{\Bibitem}[1]{\bibitem{#1}}
\nc{\Label}[1]{\label{#1}}
\nc{\eea}{\vspace{\undereqskip}\end{eqnarray}}
\nc{\ee}{\vspace{\undereqskip}\end{equation}}
\nc{\bdm}{\begin{displaymath}}
\nc{\edm}{\end{displaymath}}
\nc{\dpsty}{\displaystyle}
\nc{\bc}{\begin{center}}
\nc{\ec}{\end{center}}
\nc{\ba}{\begin{array}}
\nc{\ea}{\end{array}}
\nc{\bab}{\begin{abstract}}
\nc{\eab}{\end{abstract}}
\nc{\btab}{\begin{tabular}}
\nc{\etab}{\end{tabular}}
\nc{\bit}{\begin{itemize}}
\nc{\eit}{\end{itemize}}
\nc{\ben}{\begin{enumerate}}
\nc{\een}{\end{enumerate}}
\nc{\bfig}{\begin{figure}}
\nc{\efig}{\end{figure}}
\nc{\arreq}{&\!=\!&}
\nc{\arrmi}{&\!-\!&}
\nc{\arrpl}{&\!+\!&}
\nc{\arrap}{&\!\!\!\approx\!\!\!&}
\nc{\non}{\nonumber\\*}
\nc{\align}{\!\!\!\!\!\!\!\!&&}

\def\lsim{\; \raise0.3ex\hbox{$<$\kern-0.75em
      \raise-1.1ex\hbox{$\sim$}}\; }
\def\gsim{\; \raise0.3ex\hbox{$>$\kern-0.75em
      \raise-1.1ex\hbox{$\sim$}}\; }
\nc{\DOT}{\hspace{-0.08in}{\bf .}\hspace{0.1in}}
\nc{\Laada}{\hbox {$\sqcap$ \kern -1em $\sqcup$}}
\nc\loota{{\scriptstyle\sqcap\kern-0.55em\hbox{$\scriptstyle\sqcup$}}}
\nc\Loota{{\sqcap\kern-0.65em\hbox{$\sqcup$}}}
\nc\laada{\Loota}
\nc{\qed}{\hskip 3em \hbox{\BOX} \vskip 2ex}

\nc{\real}{{\rm I \! R}}
\nc{\Z}{{\sf Z \!\!\! Z}}
\nc{\complex}{{\rm C\!\!\! {\sf I}\,\,}}
\def\bigid{\leavevmode\hbox{\small1\kern-3.8pt\normalsize1}}
\def\id{\leavevmode\hbox{\small1\kern-3.3pt\normalsize1}}
\nc{\slask}{\!\!\!/}
\nc{\bis}{{\prime\prime}}
\nc{\pa}{\partial}
\nc{\na}{\nabla}
\nc{\ra}{\rangle}
\nc{\la}{\langle}
\nc{\goto}{\rightarrow}
\nc{\swap}{\leftrightarrow}

\nc{\EE}[1]{ \mbox{$\cdot10^{#1}$} }
\nc{\abs}[1]{\left|#1\right|}
\nc{\at}[2]{\left.#1\right|_{#2}}
\nc{\norm}[1]{\|#1\|}
\nc{\abscut}[2]{\Abs{#1}_{\scriptscriptstyle#2}}
\nc{\vek}[1]{{\rm\bf #1}}
\nc{\integral}[2]{\int\limits_{#1}^{#2}}
\nc{\inv}[1]{\frac{1}{#1}}
\nc{\dd}[2]{{{\partial #1}\over{\partial #2}}}
\nc{\ddd}[2]{{{{\partial}^2 #1}\over{\partial {#2}^2}}}
\nc{\dddd}[3]{{{{\partial}^2 #1}\over
	{\partial #2 \partial #3}}}
\nc{\dder}[2]{{{d #1}\over{d #2}}}
\nc{\ddder}[2]{{{d^2 #1}\over{d {#2}^2}}}
\nc{\dddder}[3]{{d^2 #1}\over
	{d #2 d #3}}
\nc{\dx}[1]{d\,^{#1}x}
\nc{\dy}[1]{d\,^{#1}y}
\nc{\dz}[1]{d\,^{#1}z}
\nc{\dl}[1]{\frac{d\,^{#1}l}{(2\pi)^{#1}}}
\nc{\dk}[1]{\frac{d\,^{#1}k}{(2\pi)^{#1}}}
\nc{\dq}[1]{\frac{d\,^{#1}q}{(2\pi)^{#1}}}

\nc{\cc}{\mbox{$c.c.$ }}
\nc{\hc}{\mbox{$h.c.$ }}
\nc{\cf}{cf.\ }
\nc{\erfc}{{\rm erfc}}
\nc{\Tr}{{\rm Tr\,}}
\nc{\tr}{{\rm tr\,}}
\nc{\pol}{{\rm pol}}
\nc{\sign}{{\rm sign}}
\nc{\bfT}{{\bf T }}

\def\GeV{{\rm\ GeV}}
\def\MeV{{\rm\ MeV}}

\def\TeV{{\rm\ TeV}}

\nc{\cA}{{\cal A}}
\nc{\cB}{{\cal B}}
\nc{\cD}{{\cal D}}
\nc{\cE}{{\cal E}}
\nc{\cG}{{\cal G}}
\nc{\cH}{{\cal H}}
\nc{\cL}{{\cal L}}
\nc{\cO}{{\cal O}}
\nc{\cT}{{\cal T}}
\nc{\cN}{{\cal N}}
\nc{\rvac}[1]{|{\cal O}#1\rangle}
\nc{\lvac}[1]{\langle{\cal O}#1|}
\nc{\rvacb}[1]{|{\cal O}_\beta #1\rangle}
\nc{\lvacb}[1]{\langle{\cal O}_\beta #1 |}
\nc{\bb}{\bar{\beta}}
\nc{\bt}{\tilde{\beta}}
\nc{\ctH}{\tilde{\cal H}}
\nc{\chH}{\hat{\cal H}}
\nc{\1}{\aa}
\nc{\2}{\"{a}}
\nc{\3}{\"{o}}
\nc{\4}{\AA}
\nc{\5}{\"{A}}
\nc{\6}{\"{O}}
\nc{\al}{\alpha}
\nc{\g}{\gamma}
\nc{\Del}{\Delta}
\nc{\e}{\epsilon}
\nc{\eps}{\epsilon}
\nc{\lam}{\lambda}
\nc{\om}{\omega}
\nc{\Om}{\Omega}
\nc{\ve}{\varepsilon}
\nc{\mn}{{\mu\nu}}
\nc{\k}{\kappa}
\nc{\vp}{\varphi}

\nc{\advp}[3]{{\it  Adv.\ in\ Phys.\ }{{\bf #1} {(#2)} {#3}}}
\nc{\annp}[3]{{\it  Ann.\ Phys.\ (N.Y.)\ }{{\bf #1} {(#2)} {#3}}}
\nc{\apl}[3]{{\it  Appl. Phys. Lett. }{{\bf #1} {(#2)} {#3}}}
\nc{\apj}[3]{{\it  Ap.\ J.\ }{{\bf #1} {(#2)} {#3}}}
\nc{\apjl}[3]{{\it  Ap.\ J.\ Lett.\ }{{\bf #1} {(#2)} {#3}}}
\nc{\app}[3]{{\it Astropart.\ Phys.\ }{{\bf #1} {(#2)} {#3}}}
\nc{\cmp}[3]{{\it  Comm.\ Math.\ Phys.\ }{{ \bf #1} {(#2)} {#3}}}
\nc{\cqg}[3]{{\it  Class.\ Quant.\ Grav.\ }{{\bf #1} {(#2)} {#3}}}
\nc{\epl}[3]{{\it  Europhys.\ Lett.\ }{{\bf #1} {(#2)} {#3}}}
\nc{\ijmp}[3]{{\it Int.\ J.\ Mod.\ Phys.\ }{{\bf #1} {(#2)} {#3}}}
\nc{\ijtp}[3]{{\it Int.\ J.\ Theor.\ Phys.\ }{{\bf #1} {(#2)} {#3}}}
\nc{\jmp}[3]{{\it  J.\ Math.\ Phys.\ }{{ \bf #1} {(#2)} {#3}}}
\nc{\jpa}[3]{{\it  J.\ Phys.\ A\ }{{\bf #1} {(#2)} {#3}}}
\nc{\jpc}[3]{{\it  J.\ Phys.\ C\ }{{\bf #1} {(#2)} {#3}}}
\nc{\jap}[3]{{\it J.\ Appl.\ Phys.\ }{{\bf #1} {(#2)} {#3}}}
\nc{\jpsj}[3]{{\it J.\ Phys.\ Soc.\ Japan\ }{{\bf #1} {(#2)} {#3}}}
\nc{\lmp}[3]{{\it Lett.\ Math.\ Phys.\ }{{\bf #1} {(#2)} {#3}}}
\nc{\mpl}[3]{{\it  Mod.\ Phys.\ Lett.\ }{{\bf #1} {(#2)} {#3}}}
\nc{\ncim}[3]{{\it  Nuov.\ Cim.\ }{{\bf #1} {(#2)} {#3}}}
\nc{\np}[3]{{\it  Nucl.\ Phys.\ }{{\bf #1} {(#2)} {#3}}}
\nc{\npps}[3]{{\it  Nucl.\ Phys.\ Proc.\ Suppl.\ }{{\bf #1} {(#2)} {#3}}}
\nc{\pr}[3]{{\it Phys.\ Rev.\ }{{\bf #1} {(#2)} {#3}}}
\nc{\pra}[3]{{\it  Phys.\ Rev.\ A\ }{{\bf #1} {(#2)} {#3}}}
\nc{\prb}[3]{{\it  Phys.\ Rev.\ B\ }{{{\bf #1} {(#2)} {#3}}}}
\nc{\prc}[3]{{\it  Phys.\ Rev.\ C\ }{{\bf #1} {(#2)} {#3}}}
\nc{\prd}[3]{{\it  Phys.\ Rev.\ D\ }{{\bf #1} {(#2)} {#3}}}
\nc{\prl}[3]{{\it Phys.\ Rev.\ Lett.\ }{{\bf #1} {(#2)} {#3}}}
\nc{\pl}[3]{{\it  Phys.\ Lett.\ }{{\bf #1} {(#2)} {#3}}}
\nc{\prep}[3]{{\it Phys.\ Rep.\ }{{\bf #1} {(#2)} {#3}}}
\nc{\prsl}[3]{{\it Proc.\ R.\ Soc.\ London\ }{{\bf #1} {(#2)} {#3}}}
\nc{\ptp}[3]{{\it  Prog.\ Theor.\ Phys.\ }{{\bf #1} {(#2)} {#3}}}
\nc{\ptps}[3]{{\it  Prog\ Theor.\ Phys.\ suppl.\ }{{\bf #1} {(#2)} {#3}}}
\nc{\physa}[3]{{\it  Physica\ A\ }{{\bf #1} {(#2)} {#3}}}
\nc{\physb}[3]{{\it  Physica\ B\ }{{\bf #1} {(#2)} {#3}}}
\nc{\phys}[3]{{\it Physica\ }{{\bf #1} {(#2)} {#3}}}
\nc{\rmp}[3]{{\it  Rev.\ Mod.\ Phys.\ }{{\bf #1} {(#2)} {#3}}}
\nc{\rpp}[3]{{\it Rep.\ Prog.\ Phys.\ }{{\bf #1} {(#2)} {#3}}}
\nc{\sjnp}[3]{{\it Sov.\ J.\ Nucl.\ Phys.\ }{{\bf #1} {(#2)} {#3}}}
\nc{\spjetp}[3]{{\it Sov.\ Phys.\ JETP\ }{{\bf #1} {(#2)} {#3}}}
\nc{\yf}[3]{{\it Yad.\ Fiz.\ }{{\bf #1} {(#2)} {#3}}}
\nc{\zetp}[3]{{\it Zh.\ Eksp.\ Teor.\ Fiz.\  }{{\bf #1}  {(#2)} {#3}}}
\nc{\zp}[3]{{\it Z.\ Phys.\ }{{\bf #1} {(#2)} {#3}}}
\nc{\ibid}[3]{{\sl ibid.\ }{{\bf #1} {#2} {#3}}}
\nc{\rf}[1]{(\ref{#1})}
\nc{\nn}{\nonumber \\*}
\nc{\bfB}{\bf{B}}
\nc{\bfv}{\bf{v}}
\nc{\bfx}{\bf{x}}
\nc{\bfy}{\bf{y}}
\nc{\vx}{\vec{x}}
\nc{\vy}{\vec{y}}
\nc{\oB}{\overline{B}}
\nc{\oI}{\overline{I}}
\nc{\oR}{\overline{R}}
\nc{\rar}{\rightarrow}
\nc{\ti}{\times}
\nc{\slsh}{\hskip-5pt/}
\nc{\sm}{Standard~Model~}
\nc{\MP}{M_{\rm Pl}}
\nc{\tp}{t_{\rm Pl}}
\nc{\ave}{\bar{E}}

\nc{\eff}{{\rm eff}}
\nc{\kk}{\vek{k}}
\nc{\pp}{{\rm p}}
\nc{\ga}{g_{a\gamma}}
\nc{\vv}{\\}
\nc{\eee}{{\bf E}}
\nc{\bbb}{{\bf B}}
\nc{\qcd}{T_{\rm QCD}}
\nc{\G}{\rm \ G}
\def\vec#1{{\bf #1}}
\def\lae{\;^{<}_{\sim} \;} \def\gae{\; ^{>}_{\sim} \;} 

\def\udd{u^{c}d^{c}d^{c}}

\begin{document}
{\title{\vskip-2truecm{\hfill {{\small \\
	\hfill \\
	}}\vskip 1truecm}
{\LARGE Symmetry Non-Restoration via Order $10^{-10}$ B and L Asymmetries}}
%\vspace{1.2cm}
{\author{
{\sc  John McDonald$^{1}$}\\
{\sl\small Department of Physics and Astronomy,
University of Glasgow, Glasgow G12 8QQ, SCOTLAND}
}
\maketitle
%\vspace{1cm}
%\newpage
\begin{abstract}
\noindent

              It has been observed that a large lepton asymmetry, $n_{L}/s \approx 0.1$, 
can prevent the Higgs expectation value from going to zero at high temperature, resulting 
in the non-restoration of the $SU(2) \times U(1)$ gauge symmetry and other symmetries. 
This could allow for the elimination of domain walls and other dangerous topological defects. 
Here we show that if the reheating temperature 
after inflation, $T_{R}$, is sufficiently low then symmetry non-restoration will occur with an asymmetry 
of the order of the observed baryon asymmetry, $n_{B}/s \approx 10^{-10}$. For this to occur
$T_{R} \lae 1 \GeV$ is necessary. Remarkably, this happens to be the
reheating temperature expected in Affleck-Dine baryogenesis along a $d=6$
flat direction of the MSSM. As an example, we show that this can neatly solve the $Z_{3}$ domain wall problem of the NMSSM.

\end{abstract}
\vfil
\footnoterule
{\small $^1$mcdonald@physics.gla.ac.uk}

\thispagestyle{empty}
\newpage
\setcounter{page}{1}

\section{Introduction}

                It is well-known that gauge symmetries can evade symmetry restoration at 
high temperatures in the presence of large charge densities \cite{linde,linderev,segre,others}.
 This possibility is of interest
 for a variety of reasons, such as solving the 
domain wall and monopole problems which plague many extensions of the Standard
 Model \cite{others} or permitting a large lepton (L) asymmetry to be 
consistent with a small baryon (B) asymmetry in the presence of sphaleron processes
 \cite{segre,gelcas}. In general, symmetry non-restoration requires that the number to entropy ratio of the asymmetry 
satisfies $n/s \gae 0.1$ at the epoch of symmetry breaking. It is 
in principle possible to generate such large asymmetries, in particular via
 the Affleck-Dine mechanism in SUSY models \cite{gelcas,ad}. However, 
it is interesting to ask whether it is possible to prevent symmetry restoration via a smaller
asymmetry, in particular whether an asymmetry of the order of the observed baryon asymmetry, 
$ n_{B}/s \approx 10^{-10}$, can prevent symmetry restoration. In fact we will 
see it can, so long as the reheating temperature after inflation, $T_{R}$, is sufficiently low. The reason 
is that the dilution of the asymmetry during the matter dominated period at $T > T_{R}$ 
implies that a very small asymmetry today can correspond to a very large asymmetry 
at the time of symmetry restoration. In particular, if we require that the electroweak symmetry 
is not restored at the electroweak phase transition temperature ($T_{ew} \approx m_{W}$) then we will see that 
$T_{R} \lae 1 \GeV$ is required in order to achieve this with a $10^{-10}$ asymmetry. 
Remarkably, this reheating temperature is of the order of magnitude expected 
from Affleck-Dine baryogenesis along a $d=6$ flat direction \cite{drt,jrev}, resulting
 in a self-consistent scheme for both baryogenesis and electroweak symmetry non-restoration. 
This can also prevent the restoration of other symmetries, in particular the $Z_{3}$ 
symmetry of the next-to minimal SUSY standard model (NMSSM) \cite{nmssm}, so solving 
the NMSSM domain wall problem \cite{nmdw,jnmssm}. 

               The paper is organized as follows. In section 2 we discuss symmetry 
non-restoration in the presence of a low reheating temperature. In section 3 we discuss
 the $d=6$ Affleck-Dine baryogenesis scenario. In section 4 we discuss how the $Z_{3}$ 
domain wall problem of the NMSSM can be solved in this context. In section 5 we summarize our conclusions.

\section{Symmetry Non-restoration and Low $T_{R}$}

           For the Standard Model (SM), the Higgs expectation value in the presence of a 
non-zero neutrino density, $n_{L}$, is given by the solution of \cite{linderev}
\be{e1} \lambda v(T) [v(T)^{2} - v_{o}^{2} - 
\frac{16 a_{1} n_{L}^2}{[v(T)^{2} + a_{2} T^{2}]^{2}} 
+ a_{3}T^{2}] = 0    ~,\ee
where $v_{o}$ is the $T=0$ Higgs expectation value and the constants $a_{i}$ are given in reference \cite{linderev}. 
The condition for symmetry non-restoration by a neutrino density in the Standard Model
is then 
\be{e2} l(T) \equiv \frac{n_{L}}{s} \geq  l_{c} \equiv \frac{45}{g(T) \pi^{4}} a_{2} \sqrt{\frac{a_{3}}{a_{1}}}    ~,\ee
where $g(T)$ is the number of effectively masslsss degrees of freedom at $T$.
Numerically, $a_{2} (a_{3}/a_{1})^{1/2} \approx 10$. The corresponding Higgs expectation value is 
\be{e3} \frac{v(T)}{T}  = a_{2}^{1/3} a_{3}^{1/6} \left(\frac{l}{l_{c}}\right)^{1/3}  ~\ee
for $v(T)/T$ larger than 1. For example, with $g(T) \approx 100$ at $T \approx T_{ew}$, the condition for $SU(2)$ 
symmetry non-restoration is $ l(T) \gae  5 \times 10^{-2}$. So typically 
$l(T_{ew})$ must be greater than around $0.1$ to prevent symmetry restoration in the 
SM. If entropy is conserved, this also corresponds to present asymmetry. However, if we 
have late entropy production, a small asymmetry at present can correspond to a much larger
 asymmetry at earlier times and so could prevent symmetry restoration. To see this, suppose
 we have a matter dominated Universe at $T \lae T_{ew}$. Let $T_{R}$ be the reheating temperature,
  defined as the temperature at which the matter energy density decays away and the Universe becomes
 radiation dominated. (We assume that the energy density of the decaying matter instantly thermalizes.)
 The effect of entropy production by the decay 
of the matter is to dilute the asymmetry which exists at $T_{ew}$ \cite{dilute},
\be{e4a}  n_{L}(T_{R}) = \left(\frac{g(T_{R})}{g(T_{ew})}\right)^{2} \left( \frac{T_{R}}{T_{ew}}
\right)^{8} n_{L}(T_{ew})       ~,\ee
\be{e4b}  l(T_{R}) = \left(\frac{g(T_{R})}{g(T_{ew})}\right) \left( \frac{T_{R}}{T_{ew}}
\right)^{5} l(T_{ew})       ~.\ee
For $T < T_{R}$, $l(T_{R})$ is conserved and so corresponds to the asymmetry in the 
Universe at present. Thus if $l(T_{R}) \approx 10^{-10}$, corresponding the the present B asymmetry, then in order to have 
$l(T_{ew}) \gae 0.1$ we require that $T_{R} \lae 1.6 \times 10^{-2} T_{ew}
\approx 1-3 \GeV$. Thus a reheating temperature of less than about 1 GeV 
would ensure
that a present asymmetry of around $10^{-10}$ corresponds to a 
large enough asymmetry at $T_{ew}$ to prevent electroweak symmetry restoration. 

        So far the discussion has been based on the results for the SM in the presence of a 
non-zero neutrino density. However, similar results may be expected for other 
models, such as the minimal SUSY standard model (MSSM) with two Higgs doublets, and with other asymmetries such as a 
baryon asymmetry. To see this, it is sufficient to consider the zero temperature, 
finite density $U(1)$ model with a single Higgs scalar originally considered by Linde \cite{linde}. This is given by
\be{ld1}   {\cal L} = - \frac{1}{4} F_{\mu\nu}F^{\mu\nu} + (D_{\mu}\phi)^{\dagger}D^{\mu}\phi - V(|\phi|)
+ j^{\mu}A_{\mu}      ~,\ee
where $D_{\mu} = \partial_{\mu} + ie A_{\mu}$, $V(|\phi|) = - \mu^{2}|\phi|^{2} 
+ \lambda (\phi^{\dagger}\phi)^{2}$ and $j^{\mu}$ is the current due to the charge asymmetry. In reference 
\cite{linde} this corresponded to a fermionic current $j^{\mu} 
= \overline{\psi} \gamma^{\mu} \psi$; however, its origin is not important, as it merely acts as
 a constant souce term for the gauge field. Therefore a baryon asymmetry or an asymmetry in bosons will similarly
 lead to symmetry non-restoration. Writing down the equations of motion in the physical (unitary) gauge and
 averaging over the ground state ($<...>$) gives 
\be{ld2}   \frac{(j^{\mu})^{2}}{\sigma^{3}} = \left< \frac{\partial V}{\partial \rho} \right>    ~,\ee
where $|\phi| = (\sigma + \rho(x))/\sqrt{2} $ and $<\rho> = 0$. 
This gives for the Higgs expectation value $\sigma$ \cite{linde}
\be{ld3}  \lambda \sigma^2 = \mu^2 + \frac{(j^{\mu})^{2}}{\sigma^4}      ~.\ee
Thus for a finite charge density, $(j^{\mu})^{2} = (j^{0})^{2} > 0$, the symmetry breaking
at zero temperature is increased. The effect of finite temperature is essentially to replace 
$ \sigma^2$ in equation 8 by $\sigma^2 + \alpha T^{2}$, with $\alpha \approx 1$ \cite{linde,linderev}.

            In the case with two Higgs scalars, since the effect of $j_{\mu}$ and of finite temperature will not distinguish between $\phi_{1}$ and
 $\phi_{2}$ (assuming they have the same gauge charge), the direction along which symmetry non-restoration
 occurs will be determined purely by the potential. 
For example, suppose the potential has the form of a D-term in SUSY models, $V(\phi_{1}, \phi_{2}) = \lambda (|\phi_{1}|^{2}
- |\phi_{2}|^{2})^{2}$. (This is suggested
 by the two Higgs doublets of the MSSM.) In this case the symmetry
 breaking direction will be such that $|\phi_{1}| 
= |\phi_{2}|$. Choosing $<\phi_{1}> = <\phi_{2}>$, the model can be reformulated in terms 
of $\phi_{a} = (\phi_{1} + \phi_{2})/\sqrt{2}$ and $\phi_{b} = (-\phi_{1} + \phi_{2})/\sqrt{2}$. 
Since $<\phi_{b}> = 0$, $\phi_{b}$ and its derivatives can be set to zero in ${\cal L}$ and the model reduces to one with a single Higgs field, ${\cal L}(\phi_{1},\phi_{2},A_{\mu}) \rightarrow 
{\cal L}(\phi_{a}, A_{\mu})$. The condition 
for symmetry non-restoration at finite density and temperature along the $\phi_{a}$ direction will therefore 
be the same as in the single Higgs model, equation 7. In addition, $<\phi_{a}> \neq 0$ implies that 
$\phi_{1}$ and $\phi_{2}$ will both gain expectation values. Thus we expect in general 
that the single Higgs symmetry 
non-restoration condition will be approximately
correct for models with two or more Higgs and that both Higgs will
 typically gain expectation values due to finite density symmetry breaking. 

\section{d=6 Affleck-Dine Baryogenesis and Symmetry Non-restoration}

                 Up to now we have not commented the origin of the late entropy release or of 
the asymmetries themselves. In fact, the requirement that $T_{R} \lae 1 \GeV$ suggests a remarkably
 self-consistent scenario. This is Affleck-Dine baryogenesis along a $d=6$ flat direction of the MSSM 
scalar potential \cite{drt,jrev}. In R-parity conserving models, the baryon and/or lepton asymmetry can
 originate along $d=4,6,8...$ flat directions of the MSSM scalar potential, where $d$ is the dimension of the 
non-renormalizable term which breaks B and lifts the flat direction potential.
The corresponding potential in the early Universe is given by 
\be{ld4} V(\Phi) \approx (m^{2} - c H^{2})|\Phi|^{2} 
+ \frac{\lambda^{2}|\Phi|^{2(d-1)}
}{M_{*}^{2(d-3)}} + \left( \frac{A_{\lambda} 
\lambda \Phi^{d}}{d M_{*}^{d-3}} + h.c.\right)    ~,\ee
where $H$ is the expansion rate of the Universe, $c H^{2}$ is the order $H^2$ correction to the 
scalar mass due to the energy density of the early Universe \cite{h2} (with $c$ positive 
for Affleck-Dine baryogenesis and typically of order 1) 
and the natural scale of the non-renormalizable terms is $M_{*}$,
 where $M_{*} = M_{Pl}/8 \pi$ is the supergravity (SUGRA) mass scale. The baryon
 asymmetry forms when the Affleck-Dine scalar begins to oscillate coherently about zero, which happens
at $H \approx m$, where $m \approx 100 \GeV$ is the conventional SUSY breaking scalar mass term. 
Assuming that the CP violating phase takes its natural value, $\delta_{CP} \approx 1$, the reheating temperature is then 
{\it fixed} by the dimension of the flat direction and the magnitude of the observed baryon asymmetry. 
$d=4$ directions require 
a reheating temperature $T_{R} \approx 10^{8} \GeV$ in order to account for the observed 
B asymmetry, whilst $d=6$ directions require $T_{R} \approx 1 \GeV$ \cite{jnew}. ($d \geq 8$ directions require $T_{R} \lae 1 \MeV$, which is disfavoured by nucleosynthesis.) Thus if the B asymmetry arises from $d=6$ 
Affleck-Dine baryogenesis, then the asymmetry at $T \approx T_{ew}$ would be automatically 
large enough to prevent electroweak symmetry restoration! 

             In the simplest models the matter field dominating the energy density at low temperatures 
is the coherently oscillating inflaton, although in principle it could be any weakly coupled field, since 
all that is required is that 
some non-relativistic field dominates the energy density from the time at which the 
baryon asymmetry forms. It is a challange for inflation models to achieve the low reheating temperatures 
required, but it is possible if the inflaton decays sufficiently slowly, as in the D-term inflation
 model of reference \cite{bbdti}. In addition, it has recently been shown that in many SUSY inflation models, 
non-thermal production of gravitinos requires that $T_{R} \lae 1 \TeV$ in order not to overclose
 the Universe \cite{gtr}.
Thus low reheating temperatures after inflation may be a requirement of any successful SUSY inflation model.

         Given the very low reheating temperature, one might think that 
the temperature of the electroweak phase transition would not be attained anyway, 
making symmetry non-restoration irrelevant. However, the reheating 
temperature is not the highest temperature of the thermal background; it is merely the temperature 
at which radiation comes to dominate. The temperature at a given value of $H$ during the matter
 dominated epoch at $T > T_{R}$ is given by
\be{e5}  T \approx 0.4 (M_{Pl} H T_{R}^{2})^{1/4}  ~.\ee
Since the value of $H$ at the end of inflation is typically about $10^{13} \GeV$
(fixed by the microwave background temperature fluctuations \cite{cobe}), 
for $T_{R} \approx 1 \GeV$ the radiation at the end of inflation has a temperature 
$10^{8} \GeV \gg T_{ew}$. Thus in the absence of a charge asymmetry the 
electroweak symmetry would typically be restored, even with a very low reheating 
temperature. 

\section{Solving the NMSSM $Z_{3}$ Domain Wall Problem}

              Could this symmetry non-restoration have advantages for the cosmology of SUSY models? One
 interesting application is to the next-to-minimal SUSY Standard Model (NMSSM). The NMSSM was originally motivated as a
solution to the $\mu$ problem \cite{nilles}. The MSSM requires the explicit introduction of a SUSY mass 
parameter, $\mu H_{u}H_{d}$, which has no reason to be of the order of the SUSY breaking mass terms which 
determine the electroweak scale (although we note that models which generate the 
$\mu$ term via non-minimal SUGRA terms do exist \cite{muterm}). This problem can be overcome by
 introducing an additional singlet, $N$, together with a $Z_{3}$ symmetry which prevents any explicit mass
 term for $N$ or the Higgs doublets. The superpotential is then 
\be{e6} W = W_{MSSM} + \lambda N H_{u}H_{d} + \frac{\kappa}{3} N^{3}   ~,\ee
where $W_{MSSM}$ is the MSSM superpotential. 
However, this $Z_{3}$ is broken at the electroweak phase transition, producing unacceptable 
domain walls. It is difficult to solve this problem by explicitly breaking the $Z_{3}$ symmetry, 
since in SUGRA models this usually introduces large radiative corrections, destabilizing the electroweak
 scale \cite{nmdw}. The alternative is to spontaneously break the $Z_{3}$ symmetry 
during inflation. This can be achieved if  $Z_{3}$ non-restoration is maintained from 
$T_{ew}$ up to inflation. 
We will show that this can happen for the case of $d=6$ Affleck-Dine baryogenesis 
\begin{footnote}{For an alternative approach to spontaneous $Z_{3}$ breaking, which does not involve
 a charge asymmetry, see \cite{jnmssm}.}\end{footnote}.

                We first note that an expectation value for 
{\it both} Higgs doublets is necessary to break the $Z_{3}$ symmetry, since we can always
define one of the Higgs doublets to be a singlet under the discrete symmetry (only 
$H_{u}H_{d}$ need be non-singlet). As discussed in section 2, expectation values for
both Higgs doublets are expected in the presence of a large charge density. 

                In Affleck-Dine baryogenesis, the asymmetry 
is created at $H \approx 100 \GeV$, when the Affleck-Dine scalar begins to 
oscillate coherently about the minimum of its potential. The asymmetry will be out of thermal equilibrium until
 the Affleck-Dine scalar thermalizes or decays. However, 
since symmetry non-restoration only requires an asymmetry to serve as a constant source term 
for the gauge field in the Lagrangian, the Affleck-Dine condensate can fulfil this role
even if it is not in thermal equilibrium. So for $H \lae 100 \GeV$, the electroweak 
and $Z_{3}$ symmetries will be broken by the B asymmetry. However, for $H \gae 100 \GeV$ there will be no 
B asymmetry. For these values of $H$ the $Z_{3}$ symmetry 
can be broken by the expectation value of Affleck-Dine field itself. This is slightly non-trivial, 
since in fact more than one flat direction must have an expectation value to do this.
The $d=6$ flat directions correspond to non-zero expectation values for 
the scalar field operators $(u^{c}d^{c}d^{c})^2$, $(d^{c}QL)^2$ or $(e^{c}LL)^2$. For example, 
if we considered just the 
$u^{c}d^{c}d^{c}$ direction, then by defining the $Z_{3}$ charge of $Q$ to be  
$-1/3$ and that of $H_{u}$, $H_{d}$ and $N$ to be equal to $1/3$, the $u^{c}$ and 
$d^{c}$ would be  invariant under $Z_{3}$ and so the $\udd$ flat direction scalar would not break the 
$Z_{3}$ symmetry. However, if we also have a second scalar along the $d^{c}QL$ direction, then, 
since $Q$ is not invariant under $Z_{3}$, the $Z_{3}$ symmetry will be broken up to 
and during inflation by the flat direction expectation values. 
It is quite natural for many flat direction scalars to be simultaneously
non-zero, since the negative order $H^{2}$ corrections to the mass squared terms
can occur for all the flat direction scalars.  
Thus $d=6$ Affleck-Dine baryogenesis with $T_{R} \approx 1 
\GeV$ can neatly solve the $Z_{3}$ domain wall problem of the NMSSM via a combination of finite density 
symmetry non-restoration at late times and explicit symmetry breaking by the Affleck-Dine scalars at very early times.

\section{Conclusions}

            We have shown that electroweak symmetry non-restoration can be induced by a 
baryon asymmetry of the order of that in the Universe at present, $n_{B}/s \approx 10^{-10}$, if the 
reheating temperature is sufficiently low, $T_{R} \approx 1 \GeV$.
This is naturally consistent with Affleck-Dine 
baryogenesis along a $d=6$ flat direction of the MSSM scalar potential. Although it is a challange for inflation models, such low reheating temperatures are 
possible if the inflaton is sufficiently long-lived \cite{bbdti}. In addition, recent work on the non-thermal production of gravitinos
  suggests that the reheating temperature must be less than 1 TeV in many SUSY inflation models \cite{gtr}, lending support to the idea of low reheating temperatures. 
We have also shown that $d=6$ Affleck-Dine 
baryogenesis can solve the $Z_{3}$ domain wall problem of the NMSSM by preventing $Z_{3}$ symmetry restoration 
from the electroweak phase transition temperature up to the inflationary era.

\subsection*{Acknowledgements}   This research has been supported by the PPARC.

\end{document}